\documentclass[12pt,a4paper,epsf]{article}
\usepackage{graphics}
\usepackage{amssymb,amsmath}
\usepackage[dvips]{lscape,graphicx}
\usepackage{cite}

\newcommand{\ct}{\cite}
\newcommand{\lb}{\label}

\newcommand{\bc}{\begin{center}}
\newcommand{\ec}{\end{center}}
\newcommand{\bd}{\begin{displaymath}}
\newcommand{\ed}{\end{displaymath}}
\newcommand{\be}{\begin{equation}}
\newcommand{\ee}{\end{equation}}
\newcommand{\ba}{\begin{array}}
\newcommand{\ea}{\end{array}}
\newcommand{\bea}{\begin{eqnarray}}
\newcommand{\eea}{\end{eqnarray}}
\newcommand{\bt}{\begin{tabular}}
\newcommand{\et}{\end{tabular}}

\newcommand{\ov}{\overline}
\newcommand{\bp}{\begin{picture}}
\newcommand{\ep}{\end{picture}}
\newcommand{\bfi}{\begin{figure}}
\newcommand{\efi}{\end{figure}}

\begin{document}

\hyphenation{ }

\title{\huge\bf Dyons near the Planck scale}

\author{\large Larisa Laperashvili ${}^{1,\, 2}$ \footnote{\large\, 
laper@itep.ru, laper@imsc.res.in}\,\, and C.R.~Das ${}^{1}$ \footnote{\large\, 
crdas@imsc.res.in}\\[5mm]
\itshape{${}^{1}$ The Institute of Mathematical Sciences, Chennai, India}
\\[0mm]
\itshape{${}^{2}$ The Institute of Theoretical and Experimental Physics,
 Moscow, Russia}}

\date{}

\maketitle

\vspace{1cm}

\thispagestyle{empty}

\clearpage\newpage

\thispagestyle{empty}

\begin{abstract}

In the present letter we suggest a new model of preons-dyons making 
composite quark-leptons
and bosons, described by the supersymmetric string-inspired flipped 
$E_6\times \widetilde{E_6}$ gauge group of symmetry. This approach 
predicts the 
possible extension of the Standard Model to the Family replicated gauge 
group model of type $G^{N_{fam}}$, where $N_{fam}$ is the number of families 
and $G$ is the symmetry group: $G=$ $SMG$, $SU(5)$, $SO(10)$, $E_6$, etc.
Here $E_6$ and $\widetilde{E_6}$ are non-dual and dual sectors of theory 
with hyper-electric $g$ and hyper-magnetic $\tilde g$ charges, respectively.
Starting with an idea that the most realistic model 
leading to the unification of all  fundamental 
interactions (including gravity) is the ``heterotic'' string-derived flipped 
model, we have assumed that at high energies $\mu > 10^{16}$ GeV
there exists the following chain of the flipped models:
$$ SU(3)_C\times SU(2)_L\times U(1)_Y \to
 SU(3)_C\times SU(2)_L\times U(1)_Z \times U(1)_X \to 
$$ $$ SU(5)\times U(1)_X \to  SU(5)\times U(1)_{Z1}
\times U(1)_{X1} \to SO(10) \times U(1)_{X1} \to E_6,$$
ended by the flipped $E_6$ gauge group of symmetry at the scale 
$M_{SSG}\sim 10^{18}$ GeV. Suggesting ${\rm\bf N}=1$ supersymmetric 
$E_6\times \widetilde{E_6}$ preonic model we have considered 
preons as dyons confined by hyper-magnetic strings in the region of 
energies $\mu \lesssim M_{Pl}$. 
Our model is based on the recent theory of composite non-Abelian flux tubes 
in SQCD -- analog ANO-strings. Considering the breakdown of $E_6$ 
and $\widetilde{E_6}$ at the Planck scale 
into the $SU(6)\times U(1)$ gauge group, we have shown that the 
six types of $k$-strings -- composite ${\rm\bf N}=1$ supersymmetric non-Abelian flux 
tubes -- are created by the condensation of spreons-dyons near the Planck scale
and have six fluxes quantized according to the $Z_6$ center group of
$SU(6)$: $
      \Phi_n=n\Phi_0$ $(n=\pm 1,\pm 2,\pm 3).$
These fluxes give three types of $k$-strings with tensions $T_k=kT_0$, where $k=1,2,3$, 
and produce three (and only three) generations of composite
quark-leptons and bosons giving a very specific type of ``horizontal symmetry''.
Thus, the present model predicts $N_{gen}=N_{fam}=3$.
It was shown that our preonic strings are very thin,
with radius $R_{str}\sim 10^{-18}$ GeV$^{-1}$,
and their tension $T_0$ is enormously large: $T_0\sim  10^{38}$ GeV$^2$.
It was shown that the condensation of spreons near the Planck scale 
gives the phase transition at some scales $M_{crit}$ and 
$\widetilde{M}_{crit},$ which correspond to the following breakdowns of $E_6$ 
(or $\widetilde{E_6}$) for preons:
$   E_6 \to SU(6)\times U(1)$, or 
                  $\widetilde{E}_6 \to \widetilde {SU(6)}\times 
                                        \widetilde {U(1)}.
$
We have calculated the critical values of gauge coupling constants: 
$
 \alpha^{-1}(M_{crit}) \approx 4.23$ and
$\alpha^{-1}(\widetilde{M}_{crit}) \approx 2.13.$ 
It was investigated that in our world we have quark-leptons and gauge bosons $A_{\mu}$ 
in the region of energies $\mu \lesssim M_{Pl}$, but monopolic 
``quark-leptons" and dual gauge fields $\widetilde{A}_{\mu}$ exist in the region
$\mu \gtrsim M_{Pl}$.

\end{abstract}

\clearpage\newpage

\thispagestyle{empty}

\bc
     {\Large \bf Contents:}
\ec

{\large \bf

\begin{itemize}

\item[1. ] Introduction: Superstring theory and the `flipped'\\ 
$E_6$-unification of gauge interactions.\\

\item[2. ] Supersymmetric  $E_6\times \widetilde {E_6}$ preonic model
of\\ composite quark-leptons and bosons.\\

\item[3. ] Preons are dyons confined by hyper-magnetic strings.\\

\item[4. ] The breakdown of  $E_6$  and  $\widetilde{E_6}$ groups near 
the Planck scale.\\

\item[5. ] Condensation of spreons near the Planck scale.\\

\item[6. ] ${\rm\bf N}=1$ supersymmetric non-Abelian flux tubes\\ confined  preons.\\

\item[7. ] Three generations of composite quark-leptons and bosons.\\

\item[8. ] Family replicated gauge group  $[E_6]^3$ near the Planck scale.

\end{itemize}}

\pagenumbering{arabic}

\clearpage\newpage

\setcounter{page}{1}

\begin{itemize}

\item[{\bf 1.}] Superstring theory is a paramount candidate for the ultimate theory
unifying all fundamental interactions including gravity. It was shown 
in  Refs.~\ct{1} that superstrings are free of gravitational and Yang-Mills 
anomalies if a gauge group of symmetry is $SO(32)$, or $E_8\times E_8$. 
The ``heterotic" superstring theory $E_8\times E'_8$ was suggested in \ct{1} 
as a more realistic model for unification. This ten-dimensional Yang-Mills 
theory can undergo spontaneous compactification for which $E_8$ group is 
broken to $E_6$ in four dimensions, but ${E'}_8$ group remains unbroken 
and gives ``hidden sector" of SUGRA.

In the present investigation we develop a new preonic $E_6\times \widetilde{E}_6$     
model of composite quark-leptons and bosons, in which preons are dyons confined by 
hyper-magnetic strings. Here $E_6$ and $\widetilde{E_6}$ are non-dual and dual sectors 
of theory with hyper-electric $g$ and hyper-magnetic $\tilde g$ charges, respectively.

Pati was first \ct{2} who suggested to use the strong magnetic force
to bind preons-dyons making the composite objects.
This idea has an extension in our model \ct{2a} (see also the talk \ct{6}), 
which was constructed in the light of recent 
investigations of composite non-Abelian flux tubes in SQCD \ct{3,4a,4b}.

We start with the `flipped' supersymmetric group of symmetry 
$E_6\times \widetilde{E_6}$ and show that the dual sector of this theory described
by the group $\widetilde{E_6}$ is broken in our world up to 
the Planck scale $M_{Pl}\approx 1.22\cdot 10^{19}$ GeV.
The breakdown of the dual sector gives a very specific type of the
``horizontal symmetry" predicting three generations of the Standard Model.

\item[{\bf 2.}] In Ref.~\ct{4} we have considered that only `flipped' $SU(5)$  
unifies $SU(3)_C$ and $SU(2)_W$ of the Standard Model (SM) at the GUT scale 
$M_{GUT}\sim 10^{16}$ GeV. An explanation of the discrepancy between the unification 
scale $M_{GUT}$ and string scale $M_{str}\sim 10^{18}$ GeV was given by the assumption  
that there exists a chain of extra intermediate symmetries between $M_{GUT}$ 
and $M_{Pl}$:
\be
      SU(5)\times U(1)_X \to  SU(5)\times U(1)_{Z1}
\times U(1)_{X1} \to $$ $$ SO(10) \times U(1)_{X1} \to  E_6. \lb{1}
\ee
We have considered such Higgs boson contents of the $SU(5)$ and $SO(10)$ 
gauge groups, which give the 
flipped $E_6$ final unification at the scale $\sim 10^{18}$ GeV and decreased 
running of the inversed gauge coupling constant $\alpha^{-1}$ near the Planck 
scale. Such an example, presented  by Fig.~\ref{f1}, suits the purposes of our
new model of preons.    
Here and below we consider the flipped models in which $SU(5)$ contains 
Higgs bosons $h,h^{\rm\bf c}$ and $H,H^{\rm\bf c}$ belonging to $5_h + \bar 5_h$ and 
$10_H + \ov{10}_H$ representations of $SU(5)$, respectively, also 
24-dimensional adjoint Higgs field $A$ and Higgs bosons belonging
to additional higher representations. Correspondingly, the flipped $SO(10)$ 
(coming at the superGUT scale $M_{SG}$) contains   $10_h + \ov {10}_h$ and 
$45_H + \ov{45}_H$, 45-dimensional adjoint $A$ and higher representations of 
Higgs bosons. As it was shown in \ct{4}, such Higgs boson contents lead 
to the flipped $E_6$ final unification at the supersuperGUT scale 
$M_{SSG}\sim 10^{18}$ GeV. 

Fig.~\ref{f1} presents an example of running of the inversed gauge coupling constants 
$\alpha_i^{-1}(\mu)$ ($\mu$ is the energy scale) for 
$i=1,2,3,X,Z,X1,Z1,5,10$. It was shown that at the scale $\mu = M_{GUT}$ the 
flipped $SU(5)$ undergoes the breakdown to the supersymmetric (MSSM) 
$SU(3)_C\times SU(2)_L\times U(1)_Z\times U(1)_X$ gauge group of symmetry, 
which is the supersymmetric extension of the MSSM
originated at the seesaw scale $M_{SS}\approx 10^{11}$ GeV, where heavy 
right-handed neutrinos appear. A singlet Higgs field $S$ provides the 
following breakdown to the SM (see \ct{5}):
\be SU(3)_C\times SU(2)_L\times U(1)_Z\times U(1)_X \to
    SU(3)_C\times SU(2)_L\times U(1)_Y.                            \lb{2} 
\ee
Supersymmetry extends the conventional SM beyond the scale $M_{SUSY}$. 
In our example $M_{SUSY} = 10$ TeV.

The final unification $E_6$ assumes the existence of three 27-plets of $E_6$ 
containing three generations of quarks and leptons including the right-handed 
neutrinos $N_i^{\rm\bf c}$ (here $i=1,2,3$ is the index of generations). Quarks and 
leptons of the fundamental 27 representation decompose under 
$SU(5)\times U(1)_X$ subgroup as follows:
\be
        27 \to (10,1) + \left(\bar 5, -3\right) +  \left(\bar 5,2\right) +
          (5,-2) + (1,5) + (1,0).                             \lb{3}
\ee
The first and second quantities in the brackets of Eq.~(\ref{3}) correspond to 
the $SU(5)$ representation and $U(1)_X$ charge, respectively. 
We consider charges $Q_X$ and $Q_Z$ in the units ${1}/{\sqrt{40}}$ 
and $\sqrt{3/5}$, respectively, using assignments: $Q_X=X$ and $Q_Z=Z$ \ct{4}. 

The conventional SM family which contains the doublets of left-handed quarks 
$Q$ and leptons $L$, right-handed up and down quarks $u^{\rm\bf c}$, $d^{\rm\bf c}$, 
also $e^{\rm\bf c}$,  
is assigned to the $(10,1) + (\bar 5,-3) + (1,5)$ representations 
of the flipped $SU(5)\times U(1)_X$, along with right-handed neutrino $N^{\rm\bf c}$. 
These representations decompose under
\be
SU(5)\times U(1)_X \to SU(3)_C\times SU(2)_L\times U(1)_Z\times U(1)_X.      
                                                     \lb{4a}
\ee
This decomposition for the $E_6$ 27-plet is given by Dr C.R.~Das in his talk \ct{6}.

It is necessary to notice that the flipping of our models:
\be
       d^{\rm\bf c} \leftrightarrow u^{\rm\bf c},\quad 
N^{\rm\bf c}\leftrightarrow e^{\rm\bf c}                     \lb{10}
\ee
distinguishes our `flipped $SU(5)$' from the standard Georgi-Glashow $SU(5)$.

We have the following 16 spinorial representation 
of $SO(10)$:
\be
       F(16) = F(10,1) + F\left(\bar 5,-3\right) + F(1,5).     \lb{11}
\ee
Higgs chiral superfields occupy the 10 representation of $SO(10)$:
\be 
       h(10) = h(5,-2) + h^{\rm\bf c}\left(\bar 5,2\right).            \lb{12}
\ee

\item[{\bf 3.}]
Why three generations exist in Nature ? We suggest an 
explanation considering a new preonic model of composite SM particles.
The model starts from the supersymmetric flipped $E_6\times \widetilde {E_6}$ 
gauge group of symmetry for preons.

Considering the ${\rm\bf N}=1$ supersymmetric flipped $E_6\times \widetilde {E_6}$
gauge theory for preons in $4D$-dimensional space-time, we assume that preons 
$P$ and antipreons $P^{\rm\bf c}$ are 
dyons with charges $g$ and $\tilde g$, respectively, resided
in the $4D$ hypermultiplets ${\cal P} = (P,P^{\rm\bf c})$ and ${\cal \tilde P} = 
(\tilde P, {\tilde P}^{\rm\bf c})$. Here ``$\tilde P$" designates spreons, but not 
the belonging to $\widetilde{E_6}$.

The dual sector $\widetilde{E_6}$ is broken in our world to some group $\widetilde{G}$,
and preons and spreons transform under the hyper-electric gauge group $E_6$ 
and hyper-magnetic gauge group $\widetilde{G}$ as their fundamental representations:
\be
   P,\tilde P\sim (27,N), \quad  P^{\rm\bf c},\tilde P^{\rm\bf c}\sim 
\left(\ov {27},\bar N\right),
                                         \lb{19}
\ee
where $N$ is the $N$-plet of $\widetilde{G}$ group.
We also consider scalar preons and spreons as singlets of $E_6$:
\be
   P_s,\tilde P_s\sim (1,N), \quad  P_s^{\rm\bf c},\tilde P_s^{\rm\bf c}\sim 
\left(1,\bar N\right),
                                         \lb{20}
 \ee  
which are actually necessary for the entire set of composite 
quark-leptons and bosons. This idea was suggested in Ref.~\ct{7}.

The hyper-magnetic interaction is assumed to be responsible for the formation
of $E_6$ fermions and bosons at the compositeness scale $\Lambda_s$.
The main idea of the present investigation is an assumption that
preons-dyons are confined by hyper-magnetic supersymmetric 
non-Abelian flux tubes which are a generalization of the well-known 
Abelian Abrikosov-Nielsen-Olesen (ANO)-strings for the case of the supersymmetric 
non-Abelian theory developed in Refs.~\ct{3,4a,4b}.
As a result, in the limit of infinitely narrow flux tubes (strings) we have 
the following bound states:

\begin{itemize}
\item[{\bf i.}] quark-leptons (fermions belonging to the $E_6$ fundamental representation):
\bea
            Q^a &\sim& P^{aA}(y) \left[{\cal P}\exp\left(i\tilde g
\int_x^y\widetilde{A}_{\mu}dx^{\mu}\right)\right]_A^B
                         (P_s^{\rm\bf c})_B(x) \sim 27,
                                               \lb{21}
\\
          \bar  Q_a &\sim& (P_s^{\rm\bf c})^A(y) 
\left[{\cal P}\exp\left(i\tilde g\int_x^y\widetilde{A}_{\mu}dx^{\mu}\right)\right]_A^B
                         P_{aB}(x) \sim \ov {27},
                                               \lb{22}
\eea     
where $a\in 27$-plet of $E_6$, $A, B\in N$-plet of $\widetilde{G}$, and 
$\widetilde{A}_{\mu}(x)$ are dual hyper-gluons belonging to the adjoint 
representation of $\widetilde{G}$;

\item[{\bf ii.}] ``mesons" (hyper-gluons and hyper-Higgses of $E_6$):
\bea
            M^a_b &\sim& P^{aA}(y) \left[{\cal P}\exp\left(i\tilde g
\int_x^y\widetilde{A}_{\mu}dx^{\mu}\right)\right]_A^B
                         (P^{\rm\bf c})_{bB}(x)\nonumber\\ 
&&\sim 1+78+650\quad {\rm of}\,\,E_6,
                                               \lb{23}
\\
            S &\sim& (P_s)^A(y) \left[{\cal P}\exp\left(i\tilde g
\int_x^y\widetilde{A}_{\mu}dx^{\mu}\right)\right]_A^B
                         (P_{s}^{\rm\bf c})_B(x) \sim 1,
                                               \lb{24}
\eea

\item[{\bf iii.}] ``baryons" of $\widetilde{G}$-triplet (see below):
\bea
             D_1 &\sim& \epsilon_{ABC}P^{aA'}(z)P^{bB'}(y)P^{cC'}(x)
   \left[{\cal P}\exp\left(i\tilde g\int^z_X\widetilde{A}_{\mu}dx^{\mu}\right)\right]_{A'}^A
\times
\nonumber\\ &&
 \left[{\cal P}\exp\left(i\tilde g\int^y_X\widetilde{A}_{\mu}dx^{\mu}\right)\right]_{B'}^B
 \left[{\cal P}\exp\left(i\tilde g\int^x_X\widetilde{A}_{\mu}dx^{\mu}\right)\right]_{C'}^C,
      \lb{26}
\\
             D_2 &\sim& \epsilon_{ABC}P^{aA'}(z)P^{bB'}(y)(P_s)^{C'}(x)
   \left[{\cal P}\exp\left(i\tilde g\int^z_X\widetilde{A}_{\mu}dx^{\mu}\right)\right]_{A'}^A
\times
\nonumber\\ &&
 \left[{\cal P}\exp\left(i\tilde g\int^y_X\widetilde{A}_{\mu}dx^{\mu}\right)\right]_{B'}^B
 \left[{\cal P}\exp\left(i\tilde g\int^x_X\widetilde{A}_{\mu}dx^{\mu}\right)\right]_{C'}^C,
      \lb{27}
\eea 
and their conjugate particles.
\end{itemize}

The bound states (\ref{21})--(\ref{27}) are shown
in Fig.~\ref{f3} as unclosed strings $(a)$ and ``baryonic'' configurations $(b)$.
It is easy to generalize Eqs.~(\ref{21})--(\ref{27}) for the case of 
string constructions of superpartners -- squark-sleptons, hyper-gluinos 
and hyper-higgsinos. Closed strings -- gravitons -- are presented 
in Fig.~\ref{f3}(c).
All these bound states belong to the $E_6$ representations and they are in 
fact the ${\rm\bf N}=1$ $4D$ superfields.

We assume that near the Planck scale preonic $E_6$ can be broken by
Higgses belonging to the 78-dimensional representation of $E_6$ (see Ref.~\ct{2a}):
\be
               E_6 \to SU(6)\times SU(2) \to SU(6)\times U(1),
                                                      \lb{28}
\ee
where $SU(6)\times U(1)$ is the largest relevant invariance group of the 78.

If $SU(6)\times U(1)$ group of symmetry works near the Planck scale, 
then we deal just with the theory of non-Abelian flux tubes in ${\rm\bf N}=1$
SQCD, which was developed recently in Refs.~\ct{3,4a,4b}.

Let us consider the condensation of spreons-dyons at the Planck scale.
One can combine the $Z_6$ center of $SU(6)$ with the elements 
$exp(i\pi )\in U(1)$ to get topologically stable string solutions possessing 
both windings, in $SU(6)$ and $U(1)$. Now onwards we assume the dual sector
of theory described by $\widetilde {SU(6)}\times \widetilde {U(1)}$,
which is responsible for hyper-magnetic fluxes. Then, 
according to the results obtained in Refs.~\ct{3,4a,4b}, we have 
a nontrivial homotopy group:
\be 
            \pi_1\left(\frac{SU(6)\times U(1)}{Z_6}\right) \neq 0, \lb{29}
\ee
and flux lines form topologically non-trivial $Z_6$ strings.

Besides $SU(6)$ and $U(1)$ gauge bosons, the model contains six scalar fields 
charged with respect to $U(1)$ and belong to the 6-plet of $SU(6)$.
Considering scalar fields of spreons  
\be
        \tilde P=\left\{\phi^{aA}\right\}, \lb{30}
\ee
which have indices $a$ of  $SU(6)$ and $A$ of $\widetilde {SU(6)}$ 
fundamental multiplets, we construct condensation of spreons in vacuum:
\be
       {\tilde P}_{vac} = 
\left\langle\tilde P^{aA}\right\rangle = v\cdot{\rm diag}(1,1,..,1),\quad
                               a,A=1,..,6. \lb{31}
\ee
Now we give the solution for the preonic $\bf N$=1 supersymmetric non-Abelian
flux tubes based on theory \ct{3,4a,4b}. Dual symmetry included in our model 
slightly modifies theory \ct{3,4a,4b} by consideration of the Zwanziger formalism \ct{13,14}.

\item[{\bf 4.}]
As it was shown in Refs.~\ct{13,14}, the aim to describe symmetrically 
non-dual and dual Abelian fields $A_{\mu}$ and ${\widetilde A}_{\mu}$, covariantly 
interacting with
electric $j_{\mu}^{(e)}$ and magnetic $j_{\mu}^{(m)}$ currents respectively, is realized by
the following Zwanziger's action:
\be
        S^{ZW} = \int d^4x\left(-\frac 12 n^{\mu}n^{\lambda}{\sqrt{-g}}g^{\nu\rho}\right)
     \left(F_{\mu\nu}F_{\lambda\rho} + G_{\mu\nu}F_{\lambda\rho} + iF_{\mu\nu}
          {\widetilde G}_{\lambda\rho} 
            - iG_{\mu\nu}{\widetilde F}_{\lambda\rho}\right),         \lb{a1}
\ee
where
\be
F_{\mu\nu} = \partial_{\mu}A_{\nu} - \partial_{\nu}A_{\mu},    \lb{a2}
\ee
\be
G_{\mu\nu} = \partial_{\mu}{\widetilde A}_{\nu} - \partial_{\nu}{\widetilde A}_{\mu},    \lb{a3}
\ee
and
\be {\widetilde F}_{\mu\nu} = \frac 12 \epsilon_{\mu\nu\rho\sigma} F_{\rho\sigma}. \lb{a4}
\ee 
In Eq.~(\ref{a1}) a constant unit vector $n^{\mu}$ denotes the direction of Dirac strings,
which are frozen in time and parallel in space. 

The generalized Zwanziger formalism for non-Abelian gauge theories was suggested 
in Refs.~\ct{14,12} (see also references therein), and we consider the following 
Zwanziger-type action:
\bea S^{NAZW} &=& - \frac 1{4\pi} \int {\cal D}{\xi^{\mu}(s)}ds 
\left\{ Tr\left[\left(\dot {\xi}^{\mu}(s)
F_{\mu\nu}(x)\right)\left(\dot {\xi}^{\lambda}(s)F_{\lambda}^{\nu}(x)\right)\right]
{\dot \xi}^{-2} \right.
\nonumber\\ &&+
Tr\left[\left(\dot {\xi}^{\mu}(s)
G_{\mu\nu}(x)\right)\left(\dot {\xi}^{\lambda}(s)G_{\lambda}^{\nu}(x)\right)\right]
{\dot \xi}^{-2} 
\nonumber\\ &&+
 Tr\left[\left(\dot {\xi}^{\mu}(s)
F_{\mu\nu}(x)\right)\left(\dot {\xi}^{\lambda}(s)
{\widetilde G}_{\lambda}^{\nu}(x)\right)\right]{\dot \xi}^{-2} 
\nonumber\\ &&+
Tr\left.\left[\left(\dot {\xi}^{\mu}(s)
G_{\mu\nu}(x)\right)\left(\dot {\xi}^{\lambda}(s)
{\widetilde F}_{\lambda}^{\nu}(x)\right)\right]{\dot \xi}^{-2}\right\},
\lb{a5}      
\eea 
where $\xi^{\mu}(s)$ represents an arbitrary disposition of string in $D$-dimensional
space-time ($D=4$ in our case). The functional integral  $\int {\cal D}{\xi^{\mu}}(s)$
introduces the sum over the string's shape. The ``tilde'' denotes ``dual'', but 
in this non-Abelian case it is not a simple ``Hodge star duality'' given by Eq.~(\ref{a4}) 
(see \ct{12,14}). 

In Eq.~(\ref{a5}) we have:
\be F_{\mu\nu} = \partial_{\nu}A_{\mu}(x) - \partial_{\mu}A_{\nu}(x) + i g \left[ A_{\mu}(x), 
A_{\nu}(x) \right],     \lb{a6}\ee 
and
\be G_{\mu\nu}(x) = \partial_{\nu}{\widetilde A}_{\mu}(x) -
\partial_{\mu}{\widetilde A}_{\nu}(x) + i{\tilde g} \left[ {\widetilde A}_{\mu}(x),
{\widetilde A}_{\nu}(x) \right].\lb{a7}\ee 
Here and below we do not consider the ``absorption" of the coupling constants
$g$ and $\tilde g$ by vector potentials $A_{\mu}$ and ${\widetilde A}_{\mu}$.

\item[{\bf 5.}]
Near the Planck scale the gauge group for preons is supersymmetric
$$
[SU(6)\times U(1)]\times  \left[\widetilde {SU(6)}\times \widetilde {U(1)}\right],$$
revealing the generalized dual symmetry (see \ct{13,14,12} and references therein).
In the non-dual sector, besides  $SU(6)$ and $U(1)$ gauge bosons, preons 
$P=(P_1,P^+_2)^T$ (Dirac fermions and  $SU(2)_R$ singlets)
and other matter fields, the model contains six scalar fields $\tilde P$ (spreons) 
charged with respect to $U(1)$ gauge group and belong to the 6-plet of  $SU(6)$.
In general, duality leads to the $6\times 6$ matrix (\ref{30}) for spreons.
Also they are $SU(2)_R$ doublets: $\tilde P=({\tilde P}_1,\ov{\tilde P}_2)$.
As in Refs.~\ct{3}, we introduce a zero charge scalar field $f$ and 
$35\times 35$ matrix of the complex adjoint scalar fields:
\be 
        f = \left\{f^{jJ}\right\} \quad \mbox{with} \quad j,J=1,2,..,35. \lb{b0}
\ee
In terms of these fields the action of ${\bf N}$=2 supersymmetric theory can be
deformed by the breaking to $\bf N$=1 terms containing mass 
parameters $\mu_1$ and  $\mu_2$ for the fields $f^{jJ}$ and $f$. 
The action takes the following form: 
\bea
             S &=& S^{ZW} + S^{NAZW} + S^{(\mbox{matter})} 
\nonumber\\
             &&\mbox{+ other terms (topological, gauge fixing, etc.)},      \lb{b1} 
\eea

where $S^{ZW}$ and $S^{NAZW}$ are given by Eqs.~(\ref{a1}) and (\ref{a5}),
and the third term in Eq.~(\ref{b1}) is:
\bea
  S^{(\mbox{matter})} &=& \int d^4x\left(\left|D_{\mu}f^{jJ}\right|^2 + 
                    + \left|\nabla_{\mu}{\tilde P_1}\right|^2  + 
\left|\nabla_{\mu}\ov{\tilde P_2}\right|^2
                    +  \left|\partial_{\mu}f\right|^2
                   + U\left({\tilde P},f^{jJ}, f\right)\right)  
\nonumber\\
&&\mbox{+ other matters},  \lb{b2}      
\eea
where $D_{\mu}$ is the covariant derivative in the adjoint representation 
defined as 
\be 
D_{\mu} = \partial_{\mu} - ig_6\left[ A_{\mu}(x),...\right]  - 
i{\tilde g}_6\left[ {\widetilde A}_{\mu}(x),...\right],
\lb{b3}\ee 
while 
\be 
\nabla_{\mu} = \partial_{\mu} - \frac i{\sqrt{2N}} \left(g_1A_{\mu} + 
{\tilde g}_1{\widetilde A}_{\mu}\right)
        -i \left(g_6 A_{\mu}^jT^j + {\tilde g}_6 {\widetilde A}_{\mu}^JT^J\right).
\lb{b4}\ee 
In our case $N=6$, and $T^j$, $T^J$ are $SU(6)$ generators. 
The charge $g_1$ and $(\tilde g_1)$ belongs to the 
$U(1)$ and $(\widetilde{U(1)})$ gauge group respectively and $g_6$ and $(\tilde g_6)$ 
is the charge of the
$SU(6)$ and $(\widetilde{SU(6)})$ gauge group of theory respectively.

The potential $U({\tilde P},f^{jJ}, f)$ 
is a sum of various supersymmetric $D$ and $F$ terms:
\bea
     U\left({\tilde P},f^{jJ}, f\right) &=& 
\frac 1{2g_6^{*2}}\left(\frac 1{g_6^{*2}}{\left(f^{jJ}\right)}^+f^{jJ}
     + \ov{\tilde P_1}T^jT^J{\tilde P}_1 - {\tilde P}_2T^jT^J\ov{\tilde P_2}\right)^2
\nonumber\\ &&    + \frac{g_1^{*2}}{8}\left(\left|\tilde P\right|^2 - N\xi\right)^2
    + g_6^{*2}\left(\left[{\tilde P}_2T^jT^J{\tilde P}_1\right] + \mu_2 f^{jJ}\right)^2
\nonumber\\ &&
 + g_1^{*2}\left({\tilde P}_2{\tilde P}_1 + {\mu}_1 f\right)^2 
+ \frac 12 \sum_{a,A}^6\left[\left|\left(f+f^{jJ}T^jT^J\right){\tilde P}_1^{aA}\right|^2\right.
\nonumber\\ &&
+\left.\left|\left(f+f^{jJ}T^jT^J\right){\ov{\tilde P}}_{2,aA}\right|^2\right],   \lb{b5}      
\eea
where we have charges: 
\be 
   g_1^{*2} = g_1^2 + {\tilde g}_1^2 \quad  \mbox{and} \quad   
g_6^{*2} = g_6^2 + {\tilde g}_6^2,       \lb{b6}
\ee
which are a result of the dual symmetry.

In Eq.~(\ref{b5}) we have a parameter $\xi$ belonging to the Fayet-Iliopoulos $D$-term,
which does not break $\bf N$=2 supersymmetry. The breakdown 
to $\bf N$=1 supersymmetric theory is realized with the help of parameters 
$\mu_1$ and $\mu_2$ (see \ct{3}). However, the Fayet-Iliopoulos $D$-term triggers 
the spontaneous breaking of the gauge symmetry.

The action (\ref{b1})--(\ref{b6}) gives the solution for the supersymmetric 
non-Abelin flux tubes in our preonic model. This solution was 
obtained by method developed in Ref.~\ct{3}.

The vacuum expectation value (VEV) of spreons is given as 
\be
              v =\sqrt \xi \gg \Lambda_4,           \lb{32} 
\ee
where $\Lambda_4$ is the 4-dimensional scale.
This VEV is equal to
\be
             v\sim M_{Pl}\sim 10^{19}\,\,{\rm GeV},   \lb{33}    
\ee
because we assume that spreons are condensed at the Planck scale.

Non-trivial topology (\ref{29}) amounts to winding of elements of matrix 
(\ref{30}), and we obtain string solutions:
\be
 {\tilde P}_{string} = v\cdot{\rm diag}\left(e^{i\alpha(x)},e^{i\alpha(x)},..,1,1\right),
                          \quad {\rm where}\quad x\to \infty. \lb{34}
\ee
Three types of string moduli space 
\be
   \frac{SU(6)}{SU(5)\times U(1)},\quad 
\frac{SU(6)}{SU(4)\times SU(2) \times U(1)}\quad {\rm and} \quad
\frac{SU(6)}{SU(3)\times SU(3) \times U(1)}
                                                        \lb{35}
\ee
give us solutions for three types of $Z_6$-flux tubes which are 
a non-Abelian analog of ANO-strings.

\item[{\bf 6.}] Assuming  at the ends of strings the existence of the 
preon $P$ and antipreon $P^{\rm\bf c}$ with hyper-magnetic 
charges $n\tilde g$ and $-n\tilde g$, respectively, we obtain six 
types of strings having the fluxes $\Phi_n$ quantized according 
to the $Z_6$ center group of $SU(6)$:
\be
      \Phi_n = n\Phi_0, \quad n=\pm 1,\pm 2,\pm 3.
                                                       \lb{36}          
\ee
Indeed, $Z_6$ has six group elements:
\be
        Z_6 = \left\{\left.\exp\left(2\pi \frac n6 i\right)\right|
 n\,\,{\rm mod}\,\,6\right\}. 
                                   \lb{37}   
\ee
So far as  $n$ is given by modulo 6, the fluxes of tubes corresponding 
to the solutions with $n=4,5$ are equal to the fluxes with
$n=-2,-1$, respectively (see also \ct{4a}).

String tensions of these non-Abelian flux tubes also were calculated in
Refs.~\ct{3}. The minimal tension is:
\be
               T_0 = 2\pi \xi,                  \lb{38}                  
\ee
which in our preonic model is equal to:
\be
               T_0 = 2\pi v^2\sim 10^{38}\,\,{\rm GeV}^2.
                                                          \lb{39}
\ee
Such an enormously large tension
means that preonic strings have almost infinitely small 
$\alpha' \to 0$, where $\alpha'= 1/(2\pi T_0)$ is a slope of trajectories 
in string theory \ct{1}. Three types of the preonic $k$-strings have the 
following tensions:
\be            
                    T_k =kT_0,\quad {\rm where}\quad k=1,2,3. \lb{39a}
\ee
If the Fayet-Iliopoulos term $\xi$ vanishes in the supersymmetric theory
of preons, then spreon condensate vanishes too, and the theory is in the 
Coulomb phase. But for a non-vanishing $\xi$ the spreons develop their VEV, 
and the theory is in the Higgs phase. Then hyper-magnetic charges of preons
and antipreons 
are confined by six strings which are oriented in opposite directions. By
this reason, six strings have only three different tensions (\ref{39a}).

Also preonic strings are enormously thin. Indeed, the thickness of 
the flux tube depends on the mass of the dual gauge boson $\widetilde{A}_{\mu}$
acquired in the confinement phase:
\be
                m_V =gv.                         \lb{40}
\ee
As it is shown below, in the region of energies $AB$ near the Planck scale,
where spreons are condensed (see Fig.~\ref{f2}) we have:
\be
  \alpha=\frac{g^2}{4\pi}\approx 1,\quad g\approx 2\sqrt {\pi}\approx 3.5,
                                                 \lb{41}
\ee
and the thickness of preonic strings given by the radius $R_{str}$ of 
the flux tubes is very small:
\be
      R_{str}\sim \frac 1{m_V} \sim \frac 1{gv}\sim 10^{-18}
                                          \,\,{\rm GeV}^{-1}. \lb{42}
\ee
Such infinitely narrow non-Abelian supersymmetric flux tubes remind us
superstrings of Superstring theory.
Having in our preonic model  supersymmetric strings with $\alpha'\to 0$ we
obtain, according to the description \ct{1}, only massless ground states:
spin 1/2 fermions (quarks and leptons), spin 1 hyper-gluons and spin 2 
massless graviton, as well as their superpartners. The excited states 
belonging to these strings are not realized in our world as very massive: 
they have mass $M > M_{Pl}$. 

\item[{\bf 7.}]
The hyper-flavor ``horizontal" symmetry was suggested first in 
Refs.~\ct{8}.
The previous Section gives a demonstration of a very specific 
type of the ``horizontal symmetry": 
three, and only three, generations of 
fermions and bosons present in the superstring-inspired flipped
$E_6$ theory, and also in each step given by Fig.~\ref{f1} up to
the Standard Model. This number ``3" is explained by the existence of three  
values of hyper-magnetic fluxes, which bind hyper-magnetic 
charges of preons-dyons. At the ends of these preonic strings there are placed 
hyper-magnetic charges $\pm\tilde g_0$, or $\pm 2\tilde g_0$, or 
$\pm 3{\tilde g}_0$, where ${\tilde g}_0$ is the minimal hyper-magnetic charge.
Then all bound states of Fig.~3 form three generations -- three
27-plets of $E_6$ corresponding to the three different tube flux values.
We also obtain the three types of gauge bosons $A_{\mu}^i$ ($i=1,2,3$
is the index of generations) belonging to the 
$27\times \ov {27} = 1 + 78 + 650$ representations of $E_6$. 
Fig.~4 illustrates the formation of such 
hyper-gluons: Fig.~4(a) corresponds to the composite singlet,
Fig.~4(b) and Fig.~4(c)
correspond to the adjoint 78-plet and 650-plet of hyper-gluons, respectively.

Such a description predicts the Family replicated gauge group of symmetry 
$[E_6]^3$ for quark-leptons and bosons, which works near the Planck scale. 
Here the number of families is equal to the number of generations $N_g=3$. 
We assume that $[E_6]^3$ for quark-leptons
takes place in the region of energies $M_{FR} \le \mu \le M_{crit}$, where the 
scales $M_{FR}$ and $M_{crit}$ are indicated in Fig.~\ref{f2} by points $D$ 
and $A$, respectively.
 
The breakdown $[E_6]^3\to E_6$ at the scale 
$M_{FR}$ is provided by several Higgses.  The analogous mechanism 
is described in reviews \ct{9} and references therein. Indeed,
in the Family replicated gauge group we have three types of gauge bosons 
$A_{\mu}^i$, which produce linear combinations:
\be
   A_{\mu,diag }^{(i)} = C_1^{(i)} A_{\mu}^{(1st\,\,fam.)} +    
C_2^{(i)} A_{\mu}^{(2nd\,\,fam.)} + C_3^{(i)} A_{\mu}^{(3rd\,\,fam.)},
\quad i=1,2,3.
                                                         \lb{42a}
\ee
The combination: 
\be
   A_{\mu,diag } = \frac 1{\sqrt 3}\left(A_{\mu}^{(1st\,\,fam.)} +    
    A_{\mu}^{(2nd\,\,fam.)} + A_{\mu}^{(3rd\,\,fam.)}\right)
                                                         \lb{42b}
\ee
is massless one, which corresponds to the 78 adjoint representation 
of hyper-gluons of  $E_6$. Two other combinations are massive and exist 
only in the region of energies $\mu\ge M_{FR}$. 

The assumption that only $[E_6]^3$ for quark-leptons exists in Nature 
may be not valid:
we are not sure that the Family replicated gauge groups $[SO(10)]^3$,
$[SU(5)]^3$, $[SMG\times U(1)_{(B-L)}]^3$, or
$[SMG]^3$ do not survive at lower energies $\mu < M_{FR}$.
Here we have used the following notation: $SMG$ is the Standard Model group
$SU(3)_C\times SU(2)_L\times U(1)_Y$.

The values $\alpha^{-1}(M_{crit})$ and $\alpha^{-1}(\widetilde{M}_{crit})$
can be calculated by the method developed in Ref.~\ct{10}. If $SU(N)$ group
is broken by Abelian scalar particles belonging to its Cartan $U(1)^{N-1}$
subalgebra, then $SU(N)$ critical coupling constant $\alpha_N^{crit}$ 
is given by the
following expression in the one-loop approximation (see \ct{10,12}):
\be
  \alpha_N^{crit}\approx \frac N2 \sqrt{\frac{N+1}{N-1}}\alpha_{U(1)}^{crit},
                                                               \lb{43} 
\ee
where $\alpha_{U(1)}^{crit}$ is the critical coupling constant for the
Abelian $U(1)$ theory. 

Here it is necessary to distinguish $E_6$ gauge group of symmetry for 
preons from $E_6$ for quark-leptons.  
The points $A$ and $B$ of Fig.~2 respectively correspond to the
breakdowns of $[E_6]^3$ and $[\widetilde{E}_6]^3$ to the region $AB$ 
of spreon condensation. There we have the breakdown (see Ref.~\ct{2a}) 
of the preon (one family) $E_6$ (or $\widetilde{E}_6$):  
$$   E_6 \to SU(6)\times U(1) \quad \left({\rm or} \quad 
                  \widetilde{E}_6 \to \widetilde {SU(6)}\times 
                                        \widetilde {U(1)}\right),
$$
and the scale  $M_{crit}$ (or $\widetilde{M}_{crit}$)
is the scale of breaking.

In our preonic model the group $SU(6)$ is broken by condensed Abelian 
scalar spreons-dyons belonging to the Cartan subalgebra $U(1)^5$. Then 
for the one family of preons we have:
\be
           \alpha_6^{crit}\approx 3.55\alpha_{U(1)}^{crit},   \lb{44}
\ee
according to Eq.~(\ref{43}) for $N=6$.

The behaviour of the effective fine structure constants in the vicinity of 
the phase transition point ``Coulomb-confinement" was investigated in the 
compact lattice $U(1)$ theory by Monte Carlo method \ct{11}. The following 
result was obtained:
\be 
           \alpha_{lat.U(1)}^{crit}\approx 0.20\pm 0.015,    
\quad {\tilde \alpha}_{lat.U(1)}^{crit}\approx 1.25\pm 0.010.    
                                                                 \lb{45}
\ee
The calculation of the critical coupling constants in the Higgs scalar 
monopole (dyon) model of dual $U(1)$ theory \ct{10,10a} gave the following result:
\be 
\alpha_{U(1)}^{crit}\approx 0.21, \quad
{\tilde \alpha}_{U(1)}^{crit}\approx 1.20  \quad - \quad 
{\mbox{in the Higgs monopole model,} }                
\lb{46}
\ee and
\be 
\alpha_{U(1)}^{crit}\approx 0.19, \quad
{\tilde \alpha}_{U(1)}^{crit}\approx 1.29   \quad - \quad {\mbox{in the Higgs dyon model.}}
                          \lb{46a}
\ee
According to Eqs.~(\ref{45})--(\ref{46a}), the condensation of spreons leads 
to the following critical constants:
\be
            \alpha_6^{crit}\approx 3.55\cdot 0.2 \approx 0.71,
\quad     {\tilde \alpha}_6^{crit} = \left(\alpha_6^{crit}\right)^{-1}\approx 1.41. 
                                                   \lb{47} 
\ee
For the point $C$ shown in Fig.~\ref{f2} we have:
\be
\alpha_6^{(one\,\,fam.)}(M_{Pl})=1. \lb{fam}
\ee 
This result for the one family of preons confirms the estimate (\ref{41}).
Here we have used the Dirac relation for non-Abelian theories (see explanation
in Ref.~\ct{12}):
\be
    g\tilde g = 4\pi n, \quad n\in Z, \quad \alpha \tilde \alpha = 1. 
     \lb{48}
\ee
From the phase transition result (\ref{47}) for the one family of preons,
we obtain the following inversed critical coupling constant for $[E_6]^3$
at the phase transition point $A$ (see reviews \ct{9}):
\be
              \alpha_A^{-1}(M_{crit}) = 3{\tilde \alpha}_6^{crit}(M_{crit})
                        \approx 3\cdot 1.41\approx 4.23.          \lb{50}
\ee
At the phase transition point $B$, when we have  the breakdown of 
$[\widetilde{E}_6]^3$ into the confinement phase, the one-family value 
${\tilde \alpha}_6^{crit}(\widetilde{M}_{crit})\approx 0.71$ gives the 
following result:
\be
        \alpha_B^{-1}\left(\widetilde{M}_{crit}\right)\approx 3\cdot 0.71
                       \approx 2.13.                        \lb{51} 
\ee
The values (\ref{50}) and (\ref{51}) were used for the construction of the 
curve $AB$ presented in Fig.~\ref{f2}. The point $C$ corresponds to the Planck
scale and $\alpha^{-1}(M_{Pl})=3$, according to Eq.~(\ref{fam}).

The condensation of spreons at the Planck scale predicts the existence 
of the second minimum of the effective potential $V_{eff}(\mu)$ at the scale 
$\mu=M_{Pl}$. The behaviour of this potential and its relation with the 
Multiple Point Principle -- theory of degenerate vacua by D.L.~Bennett, 
C.D.~Froggatt and H.B.~Nielsen (see reviews \ct{9} and references therein) 
will be considered in our future investigation.

Fig.~2 shows two points $A$ and $B$ near the Planck scale. The point 
$A$ corresponds to the breakdown of $E_6$ for preons 
according to the chain (\ref{28}).
The group $E_6$ is broken in the region of energies $\mu \ge M_{crit}$ 
producing hyper-electric strings between preons.  
The point $B$ in Fig.~2 indicates the scale 
$\widetilde{M}_{crit}$ corresponding  to the breakdown 
$\widetilde {E_6} \to \widetilde {SU(6)}\times \widetilde {U(1)}$.
At the point $B$ hyper-magnetic strings are produced and 
exist in the region of 
energies $\mu \le \widetilde{M}_{crit}$ confining hyper-magnetic charges of 
preons. 
As a result, in the 
region  $\mu \le M_{crit}$ we see quark-leptons with charges $ng$ $(n\in Z)$, 
but in the region  $\mu \ge \widetilde{M}_{crit}$ monopolic ``quark-leptons" 
-- particles with dual charges $m\tilde g$ $(m\in Z)$ -- may exist.
Since $\widetilde{M}_{crit} > M_{Pl}$, monopoles are absent in our world.

In the region of energies $M_{crit}\le \mu \le \widetilde{M}_{crit}$ around the 
Planck scale, both hyper-electric and hyper-magnetic strings come to play: 
preons, quarks and monopolic ``quarks" are totally confined 
giving heavy neutral particles with mass $M\sim M_{Pl}$, but closed strings -- 
gravitons -- survive there.

The dotted curve in Fig.~2 describes the running of $\alpha^{-1}(\mu)$
for monopolic ``quark-leptons'' created by preons, which are bound by 
supersymmetric hyper-electric non-Abelian flux tubes. We assume that such 
monopoles do not exist in our world, however, they can play an essential 
role in the Universe vacuum (Cosmological Constant) \ct{15}.

\end{itemize}

\bc{\Large \bf Conclusions}
\ec

\begin{itemize}
\item[{\bf i.}] In the present paper starting with an idea that the most realistic 
model based on the superstring theory is the `flipped' $E_6$ gauge group of symmetry, 
we have assumed that at high energies $\mu > 10^{16}$ GeV
there exists the following chain of the flipped models:
$$ SU(3)_C\times SU(2)_L\times U(1)_Y \to
 SU(3)_C\times SU(2)_L\times U(1)_Z \times U(1)_X \to 
$$ $$ SU(5)\times U(1)_X \to  SU(5)\times U(1)_{Z1}
\times U(1)_{X1} \to SO(10) \times U(1)_{X1} \to E_6,$$
with the flipped $E_6$ final unification.
We have chosen such Higgs
boson contents of the $SU(5)$ and $SO(10)$ gauge groups, which give 
the flipped $E_6$ at the scale $\sim 10^{18}$ GeV and the decreased 
running of $\alpha^{-1}(\mu)$ near the Planck scale.

\item[{\bf ii.}] We have shown that the final unification $E_6$ assumes the existence 
of three 27-plets of the flipped $E_6$ gauge group of symmetry.

\item[{\bf iii.}] Suggesting ${\rm\bf N}=1$ supersymmetric $E_6\times 
\widetilde{E_6}$ preonic 
model of composite quark-leptons and bosons, we have assumed that preons are 
dyons confined by hyper-magnetic strings in the region of energies 
$\mu \lesssim M_{Pl}$. This approach is an extension of the old idea by
J.~Pati to use the strong magnetic forces which may bind preons-dyons 
in composite particles -- quark-leptons and bosons. Our 
model is based on the recent theory of composite non-Abelian flux tubes 
in SQCD \ct{3,4a,4b}.

\item[{\bf iv.}] Considering the breakdown of $E_6$ (or $\widetilde{E_6}$) 
at the Planck scale
into the $SU(6)\times U(1)$ (or $\widetilde{SU(6)}\times \widetilde{U(1)}$)
gauge group we have shown that the six types of 
$k$-strings -- ${\rm\bf N}=1$ supersymmetric non-Abelian 
flux tubes -- are created by the condensation of spreons near the Planck scale.

\item[{\bf v.}] It was shown that the six types of strings-tubes having six fluxes
quantized according to the $Z_6$ center group of
$SU(6)$:
$$
      \Phi_n = n\Phi_0, \quad n=\pm 1,\pm 2,\pm 3,
$$
create three types of $k$-strings with tensions:
$$
                     T_k = kT_0, \quad k=1,2,3,
$$
which produce three (and only three) generations of composite quark-leptons and 
bosons. We have obtained a specific type of the ``horizontal symmetry'' explaining a
flavor.

\item[{\bf vi.}] It was investigated that in the present model preonic strings 
are very thin, with radius
$$
          R_{str}\sim 10^{-18}\,\,{\rm GeV}^{-1},
$$
and their tension is enormously large:
$$  
                   T_0\sim  10^{38}\,\,{\rm GeV}^2.
$$
 
\item[{\bf vii.}] The model predicts the existence of three families of 27-plets and
also gauge bosons $A^i_{\mu}$ (with $i=1,2,3$) belonging to the 78-plets of 
$E_6$. Then near the Planck scale we have the Family 
replicated gauge group of symmetry $[E_6]^3$. In the present paper 
we have assumed that the breakdown $[E_6]^3\to E_6$ occurs near the 
Planck scale 
leading to the $E_6$ unification at the scale $\sim 10^{18}$ GeV.

\item[{\bf viii.}] We have considered that the condensation of spreons-dyons near the 
Planck scale gives the phase transitions at the scales $M_{crit}$ 
and $\widetilde{M}_{crit}$ shown in Fig.~2. 
These scales (points $A,B$ of Fig.~2) correspond 
to the breakdown
of $E_6$ and $\widetilde{E}_6$ for preons, respectively:
$$   E_6 \to SU(6)\times U(1) \quad {\rm and} \quad 
                  \widetilde{E}_6 \to \widetilde {SU(6)}\times 
                                        \widetilde {U(1)}.
$$
\item[{\bf ix.}] 
It was investigated that hyper-magnetic strings are produced and exist
at $\mu \le \widetilde{M}_{crit}$, and hyper-electric strings are created and
exist at $\mu \ge M_{crit}$.
As a result, in our world we have quark-leptons and gauge bosons $A_{\mu}$ 
in the region of energies $\mu \lesssim M_{Pl}$, but monopolic 
``quark-leptons" and dual gauge fields $\widetilde{A}_{\mu}$ exist in the region
$\mu \gtrsim M_{Pl}$ (at the trans-Planckian scales, see \ct{15}).
We have calculated the critical values of gauge coupling constants
at the scales $M_{crit}$ and $\widetilde{M}_{crit}$:
$$
 \alpha^{-1}(M_{crit}) \approx 4.23, \quad {\rm and} \quad
\alpha^{-1}\left(\widetilde{M}_{crit}\right) \approx 2.13.
$$ 

\end{itemize}

{\large \bf Acknowledgements:}
C.R.D. thanks a lot Prof.~J.~Pati for interesting discussions during the 
Conference WHEPP-9 (Bhubaneswar, India, January, 2006). 
L.L. sincerely thanks the Institute of Mathematical Sciences (Chennai, India) 
and personally Director of IMSc Prof.~R.~Balasubramanian and  
Prof.~N.D.~Hari Dass for the wonderful hospitality and financial support.
The authors also deeply thank Prof.~J.L.~Chkareuli,
Prof.~F.R.~Klinkhamer and Prof.~G.~Rajasekaran for fruitful discussions and 
advices. 

This work was supported by the Russian Foundation for Basic Research (RFBR), 
project $N^o$ 05--02--17642.

\clearpage\newpage

\bfi
\centering
\includegraphics[height=163mm,keepaspectratio=true,angle=-90]{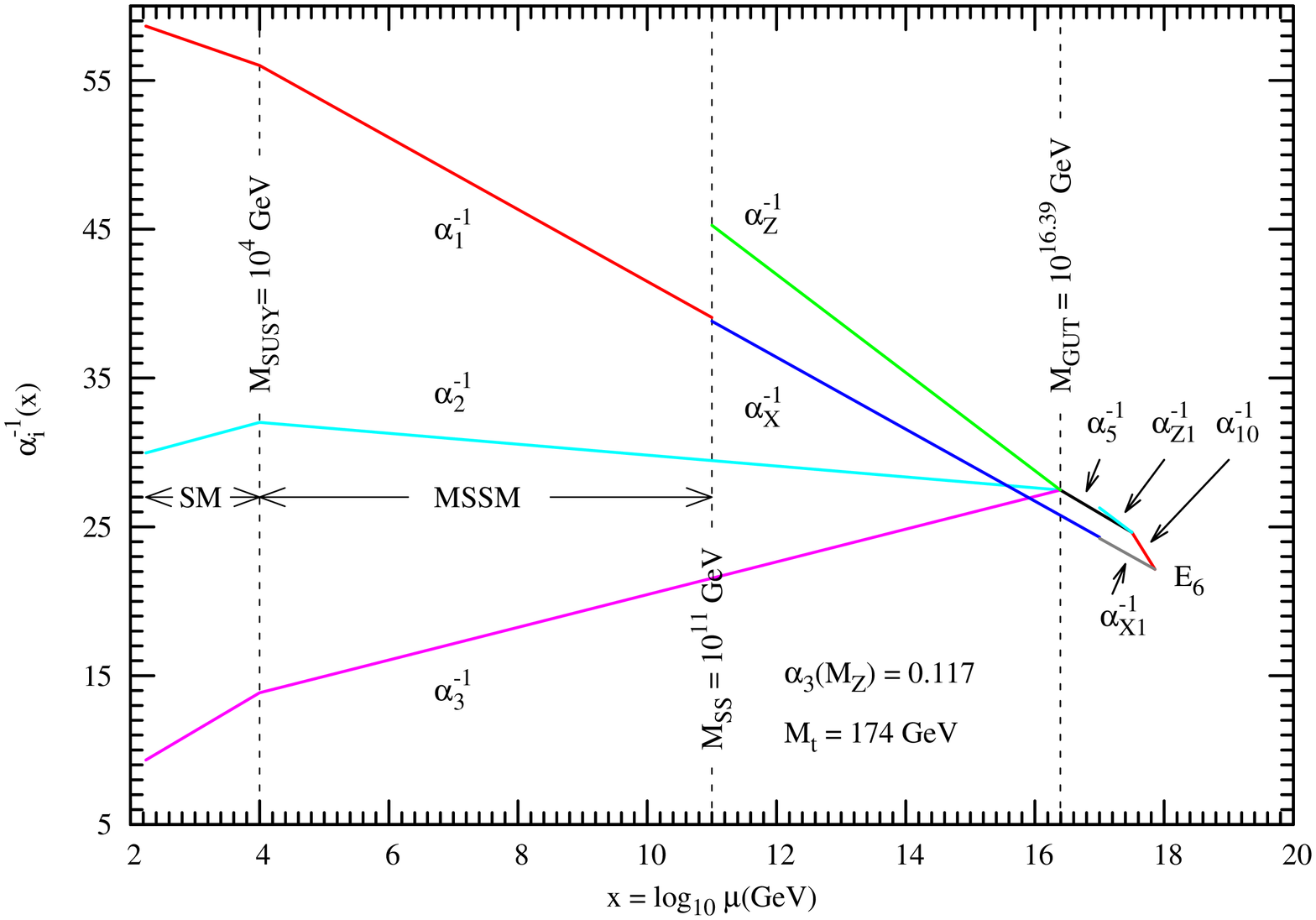}
\caption
{This figure presents the running of the inversed gauge coupling constants
$\alpha_i^{-1}(x)$ for $i=1,2,3,X,Z,X1,Z1,5,10$ of the chain
$ SU(3)_C\times SU(2)_L\times U(1)_Y \to
 SU(3)_C\times SU(2)_L\times U(1)_Z \times U(1)_X \to 
SU(5)\times U(1)_X \to  SU(5)\times U(1)_{Z1}
\times U(1)_{X1} \to SO(10) \times U(1)_{X1} \to E_6
$ corresponding to the breakdown of the flipped $SU(5)$ to the supersymmetric 
(MSSM) $SU(3)_C\times SU(2)_L\times U(1)_Z\times U(1)_X$ gauge group of symmetry
with Higgs bosons belonging to the $5_h + \bar 5_h$, $10_H + \ov {10}_H$,  
24-dimensional adjoint $A$ and higher representations of the flipped $SU(5)$.
The final unification is $E_6$ at the supersuperGUT scale $ M_{SSG}\approx 
10^{18}$ GeV with $\alpha^{-1}(M_{SSG})\approx 22$ for $M_{SUSY}=10$ TeV and 
seesaw scale $M_{SS}=10^{11}$ GeV.}
\lb{f1}
\efi

\clearpage\newpage
\bfi
\centering
\includegraphics[height=163mm,keepaspectratio=true,angle=-90]{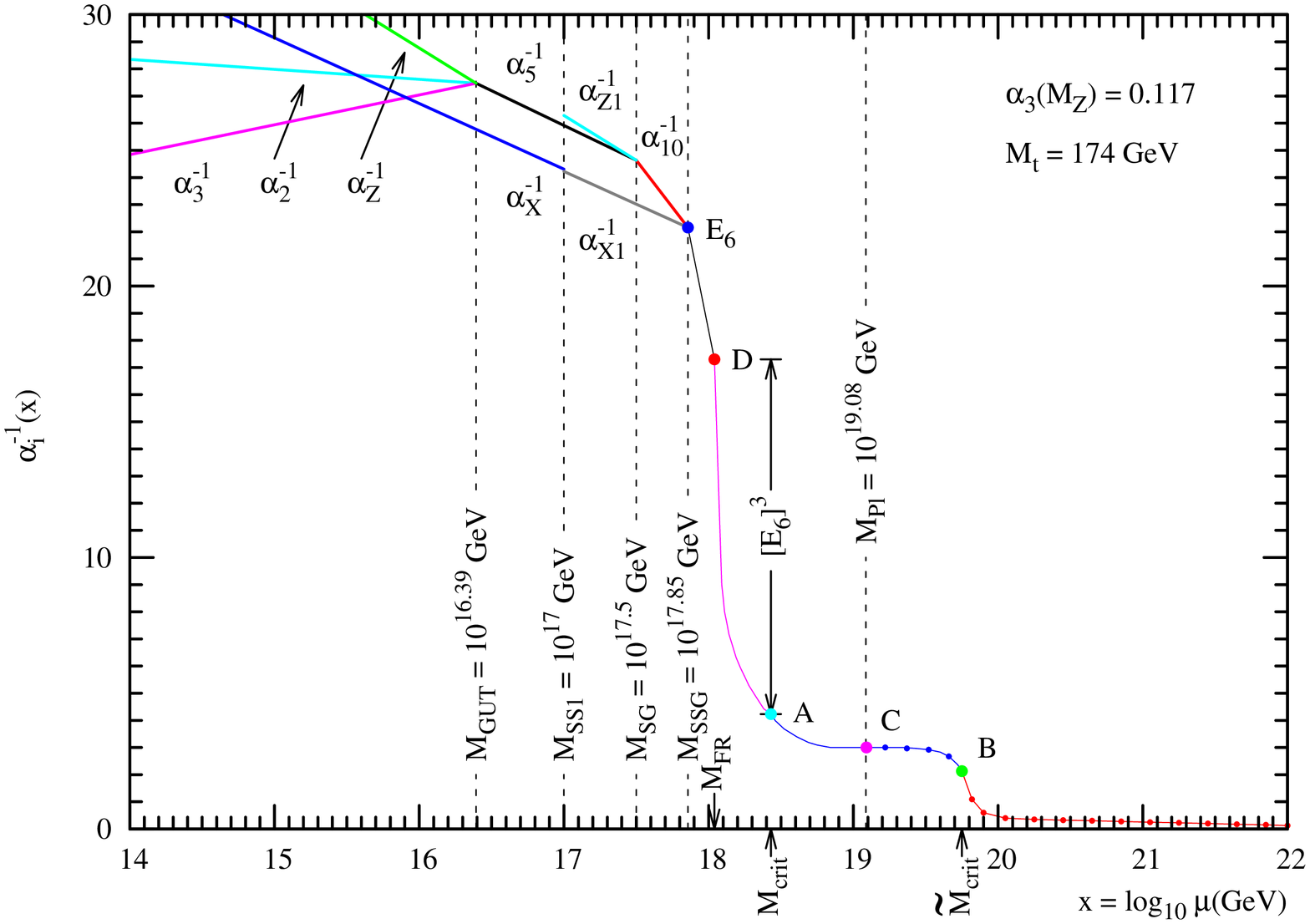}
\caption
{The figure illustrates a qualitative description of the running of $\alpha^{-1}(x)$ 
near the Planck scale predicted by the present preonic model. The region 
$AD$ corresponds to the Family replicated gauge group of symmetry $[E_6]^3$,
which comes at the scale $M_{FR}$ given by point $D$.
The point $A$ at the scale of energy $\mu=M_{crit}$ shows that hyper-electric
preonic strings exist for $\mu \ge M_{crit}$. The point $B$ corresponds to 
the scale $\mu = \widetilde{M}_{crit}$ and indicates that hyper-magnetic 
preonic strings exist for  $\mu\le \widetilde{M}_{crit}$. The curve $AB$
corresponds to the region of energies, where spreons are condensed near the
Planck scale giving both, hyper-electric and hyper-magnetic, preonic strings.
For  $\mu \ge \widetilde{M}_{crit}$ we have the running of $\alpha^{-1}(x)$
for monopolic ``quark-leptons". The point $C$ corresponds to the Planck scale
and gives $\alpha^{-1}(M_{Pl})=3$.}
\lb{f2}
\efi

\clearpage\newpage

\bfi
\centering
\includegraphics[height=62mm,keepaspectratio=true]{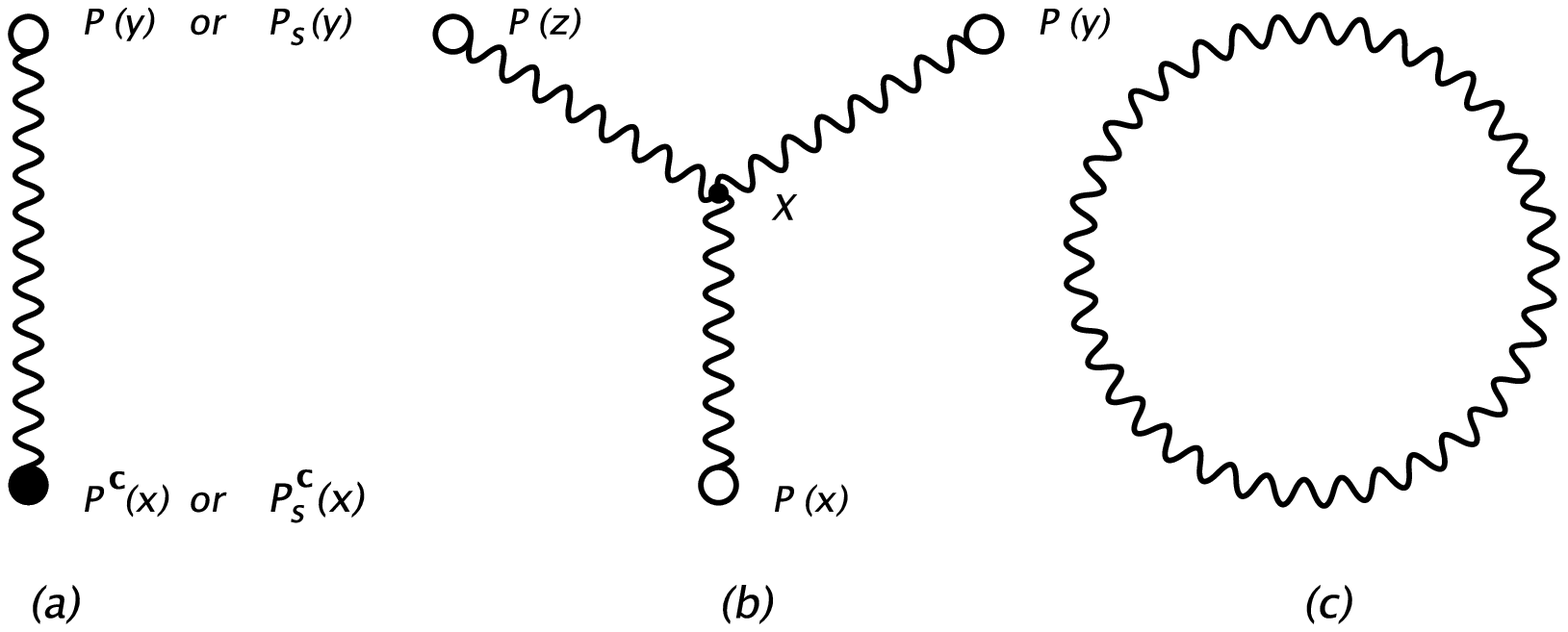}
\caption{Preons are bound by hyper-magnetic strings: (a,b) correspond to
the string configurations of composite particles belonging to the
27-plet of $E_6$ gauge group of symmetry; (c) represents a closed string 
describing a graviton.}
\lb{f3}
\efi

\clearpage\newpage
\bfi
\centering
\includegraphics[height=59mm,keepaspectratio=true]{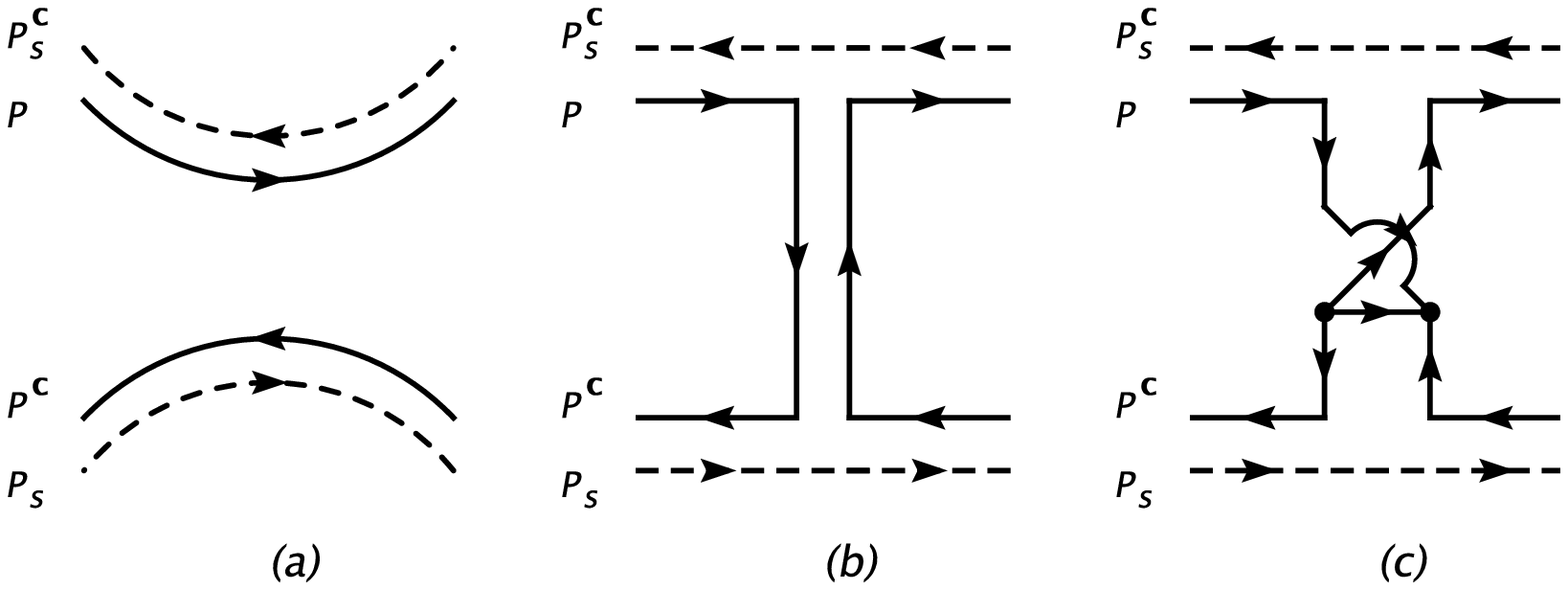}
\caption{Vector gauge bosons belonging to the 1 + 78 + 650 representations of $E_6$.
Gluons are composite objects created 
by fermionic preons $P,P^{\rm\bf c}$: (a) correponds to the singlet; (b) 
is the 78-plet and (c) is the 650-plet of $E_6$.}

\lb{preon4-talk.ps}
\efi


\begin{thebibliography}{99}
\bibitem{1}
J.~Schwarz, Phys.Rept. {\bf 89}, 223 (1982); M.B.~Green, Surv.High.En.Phys. 
{\bf 3}, 127 (1984); M.B.~Green and J.H.~Schwarz, Phys.Lett. {\bf B149}, 117 (1984); ibid., 
{\bf B151}, 21 (1985); D.J.~Gross, J.A.~Harvey, E.~Martinec and 
R.~Rohm, Phys.Rev.Lett. {\bf 54}, 
502 (1985); Nucl.Phys. {\bf B256}, 253 (1985); ibid., {\bf B267}, 75 (1986). 
\bibitem{2}
J.C.~Pati, Phys.Lett. {\bf B98}, 40 (1981);
J.C.~Pati, A.~Salam and J.~Strathdee, Nucl.Phys. {\bf B185}, 416 (1981).
\bibitem{2a}
C.R.~Das and L.V.~Laperashvili,
Phys.Rev. {\bf D74}, 035007 (2006); arXiv: hep-ph/0605161. 
\bibitem{6}
C.R.~Das and L.V.~Laperashvili, {\it Composite model of quark-leptons and duality},
a talk prepared for the 17th
DAE-BRNS High Energy Physics Symposium, IIT, Kharagpur, India, 11-15 December, 2006; arXiv: hep-ph/0606042.
\bibitem{3}
M.~Shifman and A.~Yung, Phys.Rev. {\bf D70}, 025013 (2004), arXiv: hep-th/0312257;
ibid., {\bf D70}, 045004 (2004), arXiv: hep-th/0403149; 
ibid., {\bf D72}, 085017 (2005), arXiv: hep-th/0501211;
A.~Gorsky, M.~Shifman and A.~Yung, Phys.Rev. {\bf D71}, 045010 (2005), 
arXiv: hep-th/0412082;
V.~Markov, A.~Marshakov and A.~Yung, Nucl.Phys. {\bf B709}, 267 (2005),
arXiv: hep-th/0408235;
R.~Auzzi, M.~Shifman and A.~Yung, Phys.Rev. {\bf D73}, 105012 (2006), arXiv: hep-th/0511150.
\bibitem{4a}
M.A.C.~Kneipp, Phys.Rev. {\bf D69}, 045007 (2004); arXiv: hep-th/0308086.
\bibitem{4b}
Y.~Isozumi, M.~Nitta, K.~Ohashi and N.~Sakai, Phys.Rev. {\bf D71}, 065018 (2005), 
arXiv: hep-th/0405129;
M.~Eto, Y.~Isozumi, M.~Nitta, K.~Ohashi and N.~Sakai,
Phys.Rev.Lett. {\bf 96}, 161601 (2006), 
arXiv: hep-th/0511088;
M.~Eto, T.~Fujimori, Y.~Isozumi, M.~Nitta, K.~Ohashi, K.~Ohta and N.~Sakai, 
Phys.Rev. {\bf D73}, 085008 (2006), arXiv: hep-th/0601181;
M.~Eto, Y.~Isozumi, M.~Nitta, K.~Ohashi and N.~Sakai, Phys.Rev. {\bf D73}, 125008 (2006), 
arXiv: hep-th/0602289.
\bibitem{4}
C.R.~Das and L.V.~Laperashvili, {\it Seesaw scales and the steps from the 
Standard Model towards superstring-inspired flipped $E_6$}; arXiv: hep-ph/0604052.
\bibitem{5}
C.R.~Das, C.D.~Froggatt, L.V.~Laperashvili and H.B.~Nielsen, Mod.Phys.Lett. 
{\bf A21}, 1151 (2006); arXiv: hep-ph/0507182.
\bibitem{7}
M.~Chaichian, J.L.~Chkareuli and A.~Kobakhidze, Phys.Rev. {\bf D66}, 095013 (2002); 
arXiv: hep-ph/0108131.
\bibitem{13}
D.~Zwanziger, Phys.Rev. {\bf 176}, 1489 (1968); ibid., Phys.Rev. {\bf D3}, 
880 (1971);
R.A.~Brandt, F.~Neri and D.~Zwanziger, Phys.Rev. {\bf D19}, 1153 (1979);
F.V.~Gubarev, M.I.~Polikarpov and V.I.~Zakharov, Phys.Lett. {\bf B438}, 
147 (1998), arXiv: hep-th/9805175; 
L.V.~Laperashvili and H.B.~Nielsen, Mod.Phys.Lett. {\bf A14}, 
2797 (1999), arXiv: hep-th/9910101.
\bibitem{14}
L.V.~Laperashvili, {\it Generalized dual symmetry of non-Abelian theories},
arXiv: hep-th/0211227;
A.~Noguchi and A.~Sugamoto, {\it Dynamical origin of duality between 
gauge theory and gravity}, in:  TSPU Vestnik {\bf 44N7}, 2004, p.59., 
arXiv: hep-th/0408045.
\bibitem{12}
C.R.~Das, L.V.~Laperashvili and H.B.~Nielsen, {\it Generalized dual symmetry of 
non-Abelian theories and the freezing of $\alpha_s$}, 
to appear in Int.J.Mod.Phys. {\bf A}, (2006); arXiv: hep-ph/0511267.
\bibitem{8}
J.L.~Chkareuli, JETP Lett. {\bf 32}, 671 (1980) [Pisma Zh.Eksp.Teor.Fiz. {\bf 32}, 684 (1980)]; 
Z.G.~Berezhiani and J.L.~Chkareuli, Sov.J.Nucl.Phys. {\bf 37}, 618 (1983) [Yad.
Fiz. {\bf 37}, 1043 (1983)].
\bibitem{9}
C.D.~Froggatt, L.V.~Laperashvili, H.B.~Nielsen and Y.~Takanishi,
{\it Family Replicated Gauge Group Models}, in: Proceedings of the
``Fifth International Conference Symmetry in Nonlinear Mathematical
Physics'', Kiev, Ukraine, 23-29 June, 2003, Ed. by A.G.~Nikitin,
V.M.~Boyko, R.O.~Popovich and I.A.~Yehorchenko (Institute of Mathematics
of NAS of Ukraine, Kiev, 2004), Vol.50, Part 2, p.737, arXiv:
hep-ph/0309129;
L.V.~Laperashvili, Phys.At.Nucl. {\bf 57}, 471 (1994) [Yad.Fiz. {\bf 57}, 501 (1994)]; 
ibid., {\bf 59}, 162 (1996) [Yad.Fiz. {\bf 59}, 172 (1996)].
\bibitem{10}
L.V.~Laperashvili, H.B.~Nielsen and D.A.~Ryzhikh, Int.J.Mod.Phys. {\bf A16}, 
3989 (2001), arXiv: hep-th/0105275;
L.V.~Laperashvili and H.B.~Nielsen, Int.J.Mod.Phys. {\bf A16}, 2365 (2001), 
arXiv: hep-th/0010260.
\bibitem{10a}
C.R.~Das and L.V.~Laperashvili, {\it Phase transition in the Higgs model of scalar dyons},
to appear in Mod.Phys.Lett. {\bf A} (2006); arXiv: hep-ph/0511067.
\bibitem{11}
J.~Jersak, T.~Neuhaus and P.M.~Zerwas, Phys.Lett. {\bf B133}, 103 (1983); 
Nucl.Phys. {\bf B251}, 299 (1985);
J.~Jersak, T.~Neuhaus and H.~Pfeiffer, Phys.Rev. {\bf D60}, 054502 (1999), 
arXiv: hep-lat/9903034.
\bibitem{15}
C.~Froggatt, L.~Laperashvili, R.~Nevzorov and H.B.~Nielsen,
Phys.Atom.Nucl.{\bf 67}, 582 (2004) [Yad.Fiz. {\bf 67}, 601 (2004)], arXiv: hep-ph/0310127;
F.R.~Klinkhamer and G.E.~Volovik, JETP Lett. {\bf 81}, 551 (2005), 
[Pisma Zh.Eksp.Teor.Fiz. {\bf 81}, 683 (2005)]
arXiv: hep-ph/0505033. 

\end{thebibliography}
\end{document}